\documentclass[10pt,conference]{IEEEtran}
\usepackage[font=small,skip=0.5pt]{caption}
\usepackage{amsmath}
\usepackage{amssymb}
\usepackage{amsfonts}
\usepackage{amsbsy}
\usepackage{graphicx}

\usepackage{epstopdf}
\newcommand{\MeijerG}[7]{G \begin{smallmatrix} #1, #2 \\ #3, #4 \end{smallmatrix}\!\! \left[ #5 \middle\vert\begin{matrix} #6 \\ #7 \end{matrix} \right] }
\newcommand{\MeijerGG}[6]{G^{#1}_{#2}\!\! \left[ \begin{matrix} #3 \\ #4 \end{matrix} \middle\vert\begin{matrix} #5 \\ #6 \end{matrix} \right] }

\begin{document}

\title{\huge Internet of Things-Enabled Overlay Satellite-Terrestrial Networks in the Presence of Interference}

\author{Pankaj K. Sharma, Budharam Yogesh, and Deepika Gupta}
\maketitle

\begin{abstract}
In this paper, we consider an overlay satellite-terrestrial network (OSTN) where an opportunistically selected terrestrial IoT network assist primary satellite communications as well as access the spectrum for its own communications in the presence of combined interference from extra-terrestrial and terrestrial sources. Hereby, a power domain multiplexing is adopted by the IoT network by splitting its power appropriately among the satellite and IoT signals. Relying upon an amplify-and-forward (AF)-based opportunistic IoT network selection strategy that minimizes the outage probability (OP) of satellite network, we derive the closed-form lower bound OP expressions for both the satellite and IoT networks. We further derive the corresponding asymptotic OP expressions to examine the achievable diversity order of two networks. We show that the proposed OSTN with adaptive power splitting factor benefits IoT network while guaranteeing the quality of service (QoS) of satellite network. We verify the numerical results by simulations.
\end{abstract}

\section{Introduction}
Integration of relay cooperation to satellite networks has been emerged as a popular paradigm for reliable communications, especially when the satellite-terrestrial user links are severely masked \cite{evans} (i.e., in the presence of heavy clouds, physical blockages, ground user in tunnels, etc.). Consequently, the satellite-terrestrial networks (STNs) are investigated in literature by employing both the amplify-and-forward (AF) \cite{bhatna}-\cite{pku} and decode-and-forward (DF) \cite{sreng}, \cite{kangdf} relaying techniques for their performance enhancement. On the other hand, the terrestrial wireless systems have recently evolved to provide wireless connectivity to extraordinarily large number of devices pertaining to various applications, commonly known as internet of things (IoT) \cite{iot}. Intuitively, these billions of IoT devices increase manifold the demand for spectrum resources in upcoming years. To this end, the cognitive radio provides a viable solution based on sharing the licensed spectrum of a primary network with IoT devices for secondary communications. Most popular models of cognitive radio are the underlay and overlay (please refer  \cite{manna}). In underlay model, the transmit power of secondary devices is strictly constrained to safeguard the primary network from harmful interference. On the contrary, in overlay model, the secondary devices cooperatively assist primary communications alongside their own secondary communications based on a less restrictive power splitting approach. In view of rapidly growing IoT applications, an overlay satellite-terrestrial network (OSTN) is of tremendous interest where the primary satellite spectrum (e.g., direct-to-home television bands, etc.) can be shared with IoT devices. Herein, the IoT devices not only can access the primary spectrum for their own communications, but can also incentivize the satellite network through cooperation. While the underlay STNs have been investigated in \cite{sksharma}-\cite{guo}, the OSTN have been analyzed in \cite{pkss}. Note that, in STNs, the co-channel interference originating from both extra-terrestrial (i.e., satellite interferers) and terrestrial (i.e., ground interferers) sources is inevitable. Consequently, the works in \cite{kangdf}, \cite{kangc}, \cite{pkueucnc} have analyzed the performance of STNs in the presence of terrestrial interferers only. However, the literature analyzing the performance of cognitive STNs with interference is sparse. So far, to our best knowledge, the performance analysis of OSTNs taking into account jointly the extra-terrestrial and terrestrial interference has not been addressed in literature.   

Motivated by this, in this paper, we analyze the performance of an IoT-enabled OSTN where an opportunistically-selected IoT transmitter assists the primary satellite communications and communicates with its own receiver in the presence of interference received from combined extra-terrestrial and terrestrial sources. In particular, we derive the closed-form lower bounds on outage probability (OP) of both the satellite and IoT networks. We further carry out the asymptotic OP analysis to disclose the achievable diversity order for both the satellite and IoT networks.

\section{System Description}\label{sysmod}
\setlength{\belowdisplayskip}{0pt}
\setlength{\abovedisplayskip}{0pt}
\subsection{System Model}
\begin{figure}[!t]
\centering
\includegraphics[width=2.5in]{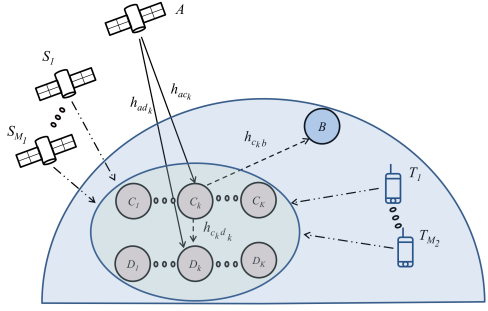}
\vspace{5pt}
\caption{OSTN with extra-terrestrial and terrestrial  interferers.}
\label{system}
\end{figure}

As shown in Fig. \ref{system}, we consider an OSTN comprising of a primary satellite transmitter ($A$)-terrestrial receiver ($B$) pair and multiple secondary IoT transmitter ($C_{k}$)-receiver ($D_{k}$) pairs, $k=1,...,K$. In addition, we consider that the multiple secondary IoT transmitter-receiver pairs are clustered where the group of IoT transmitters $\{C_k\}_{k=1}^{K}$ and the IoT receivers $\{D_k\}_{k=1}^{K}$ are inflicted by $M_1$ extra-terrestrial satellite interferers $\{S_j\}_{j=1}^{M_1}$ and $M_2$ terrestrial interferers $\{T_l\}_{l=1}^{M_2}$,  
respectively. We assume that the direct link between satellite $A$ and its receiver $B$ is masked due to severe shadowing, blocking, etc. Herein, the secondary IoT transmitters compete to utilize the primary satellite network's spectrum in lieu of opportunistically assisting the satellite-to-ground communications based on the overlay spectrum sharing principle. According to the overlay principle, a selected secondary IoT transmitter $C_{k}$ serves as a relay that splits its total transmit power $P_c$ to multiplex the received primary signal and its own secondary signal in power domain with power levels $\mu P_c$ and $(1-\mu)P_c$, respectively, where $\mu\in(0,1)$. The channels pertaining to the links $A\rightarrow C_k$, $C_k\rightarrow B$, and $C_k\rightarrow D_k$ are denoted as $h_{ac_k}$, $h_{c_kb}$, and $h_{c_kd_k}$, respectively.  Also, $\{h_{sj}\}_{j=1}^{M_{1}}$ and $\{h_{tl}\}_{l=1}^{M_{2}}$ represent the channels from $S_j$ and $T_l$ to the cluster of all IoT transmitter-receiver pairs $C_k-D_k$, $k=1,...,K$. The thermal noise at each receiver node is assumed to be additive white Gaussian noise (AWGN) with mean zero and variance $\sigma^2$.

\subsection{Channel Models}\label{srf}
\subsubsection{Main Satellite and Extra-Terrestrial Interference Channels}
The channel for main satellite links (i.e., $h_{ai_k}$, $i\in \{c,d\}$) and the extra-terrestrial interferers (i.e., $h_{sj}$) follow shadowed-Rician fading. Consequently, the probability density function (pdf) of independently and identically distributed (i.i.d.) channels $|h_{ai_k}|^{2}$ with integer-valued fading severity parameter $m_{ai}$ is given as \cite{pkss}
$f_{|h_{ai_k}|^{2}}(x)=\alpha_{i} \sum_{\kappa=0}^{m_{ai}-1}\zeta(\kappa)x^{\kappa}\textmd{e}^{-(\beta_{i}-\delta_{i})x}$,
where $\alpha_{i}=(2\flat_{ai} m_{ai}/(2\flat_{ai} m_{ai}+\Omega_{ai}))^{m_{ai}}/2\flat_{ai} $, $\beta_{i}=1/2\flat_{ai} $, and $\delta_{i}=\Omega_{ai}/(2\flat_{ai} )(2\flat_{ai} m_{ai}+\Omega_{ai})$, $\Omega_{ai}$ and $2\flat_{ai} $ are the average power of the line-of-sight and multipath components, respectively, $m_{ai}$ denotes the fading severity,  $\zeta(\kappa)=(-1)^{\kappa}(1-m_{ai})_{\kappa}\delta_{i}^{\kappa}/(\kappa!)^{2}$, and $(\cdot)_{\kappa}$ denotes the Pochhammer symbol \cite[p. xliii]{grad}. Further, the pdf of squared interferer's channel $|h_{sj}|^{2}$ can be obtained similar to the aforementioned pdf by replacing $\{\alpha_i, \beta_i, \flat_{ai}, \delta_i, m_{ai}, \Omega_{ai}\}$ as $\{\alpha_s, \beta_s, \flat_{s}, \delta_s, m_{s}, \Omega_{s}\}$ for $j=1,...,M_1$ under  i.i.d. extra-terrestrial interferers.
\subsubsection{Main Terrestrial and Terrestrial Interference Channels}

The channel for main terrestrial links (i.e., $h_{c_{k}\upsilon}$, $\upsilon\in \{b,d_{k}\}$) and terrestrial interferers follow Rayleigh fading. Hence, the pdf of i.i.d. channels $|h_{c_k\upsilon}|^2$ is given by $ f_{|h_{c_{k}\upsilon}|^2}(x)=\frac{1}{\Omega_{c\upsilon}}\textmd{e}^{-\frac{x}{\Omega_{c\upsilon}}}$, where $\upsilon\in \{b,d\}$, $k=1,...,K$ and $\Omega_{c\upsilon}$ is the average fading power. Further, the pdf of squared terrestrial interferer's channel $|h_{tl}|^2$ can be obtained similar to the above pdf by replacing $\Omega_{c\upsilon}$ with $\Omega_{t}$ for $(l=1,...,M_2)$ under i.i.d. terrestrial interferers.  

\subsection{Propagation Model}
The overall communication from satellite $A$ to terrestrial receiver $B$ takes place in two consecutive time phases with the help of a selected AF IoT transmitter relay $C_k$. Besides assisting the primary satellite communications, the IoT transmitter $C_k$ simultaneously communicates with IoT receiver $D_k$. 

In the first phase, the satellite $A$ transmits a unit energy signal $x_{a}$ to IoT transmitter $C_{k}$ with transmit power $P_{a}$, which is also received by the IoT receiver $D_{k}$. Thus, the signals received at $C_{k}$ and $D_{k}$ can be expressed as
\begin{align}\label{asrd}
y_{ai}&=\sqrt{P_{a}}h_{ai}x_{a}+{I}_{s}+{I}_{t}+n_{ai},
\end{align}
where $i\in \{c_k,d_k\}$, $I_{s}=\sum_{j=1}^{M_1}\sqrt{P_{s}}h_{sj}x_j$ and $I_{t}=\sum_{l=1}^{M_2}\sqrt{P_{t}}h_{tl}x_l$ are the interference signals received from extra-terrestrial and terrestrial interferers with corresponding transmit powers $P_s$ and $P_t$, respectively, and $n_{ai}$ is the AWGN. 

In the second phase, the IoT transmitter $C_{k}$ combines the amplified primary signal $y_{ac_{k}}$ and its own secondary signal $x_{c_{k}}$ using network coding by splitting its total power as $\mu P_c$ and $(1-\mu)P_c$ among these signals, respectively. The resulting network-coded signal can be given as 
\begin{align}\label{comb}
z_{c_{k}}=\sqrt{\mu P_{c}}\frac{y_{ac_{k}}}{\sqrt{|y_{ac_{k}}|^{2}}}+\sqrt{(1-\mu) P_{c}}x_{c_{k}},
\end{align}
where $\mu\in (0,1)$ is a power splitting factor. The IoT transmitter $C_k$ then broadcast the superposed signal $z_{c_{k}}$ which is received by the nodes $B$ and $D_k$. The received signals at $B$ and $D_k$ are given, respectively, as     
\begin{align}\label{asrd}
y_{c_{k}b}&=h_{c_{k}b}z_{c_{k}}+n_{c_{k}b},\\
\textmd{and }y_{c_{k}d_{k}}&=h_{c_{k}d_{k}}z_{c_{k}}+{I}_{s}+{I}_{t}+n_{c_{k}d_{k}},\label{asro}
\end{align}
where $I_{s}$ and $I_t$ are the same as defined previously, and $n_{c_{k}i}$, $i\in \{b,d_k\}$ is the AWGN. Thus, the signal-to-interference-plus-noise ratio (SINR) at $B$ via relay link is given by
\begin{align}\label{snrs}
\Lambda_{ac_{k}b}&=\frac{\mu\hat{\Lambda}_{ac_{k}}\Lambda_{c_{k}b}}{(1-\mu)\hat{\Lambda}_{ac_{k}}\Lambda_{c_{k}b}+\hat{\Lambda}_{ac_{k}}+\Lambda_{c_{k}b}+1},
\end{align}
where $\hat{\Lambda}_{ac_{k}}=\frac{{\Lambda}_{ac_{k}}}{W_c+1}$, $\Lambda_{ac_{k}}=\eta_{a}|h_{ac_{k}}|^{2}$, $\eta_{a}=\frac{P_{a}}{\sigma^{2}}$, $W_c\triangleq W_s+W_t$, $W_s=\sum_{j=1}^{M_1}\eta_{s}|h_{sj}|^2$, $\eta_s=\frac{{P_{s}}}{\sigma^2}$, $W_t= \sum_{l=1}^{M_2}\eta_t|h_{tl}|^2$, $\eta_t=\frac{{P_{t}}}{\sigma^2}$, $\Lambda_{c_{k}b}=\eta_{c}|h_{c_{k}b}|^{2}$, and $\eta_{c}=\frac{P_{c}}{\sigma^{2}}$. 

Moreover, from (\ref{asro}), we observe that the received signal $y_{c_{k}d_{k}}$ at $D_{k}$ contains primary satellite signal $x_a$ which can be cancelled by $D_k$ since a copy of $x_a$ is already received by it in the first phase. Hence, the equivalent SINR at IoT receiver $D_{k}$ after primary interference cancellation is given as
\begin{align}\label{secb}
\Lambda_{ac_{k}d_{k}}&=\frac{(1-\mu)\hat{\Lambda}_{c_{k}d_{k}}(\hat{\Lambda}_{ac_{k}}+1)}{\mu\hat{\Lambda}_{c_{k}d_{k}}+\hat{\Lambda}_{ac_{k}}+1},
\end{align}
where $\hat{\Lambda}_{c_{k}d_{k}}=\frac{{\Lambda}_{c_{k}d_{k}}}{W_c+1}$ and ${\Lambda}_{c_{k}d_{k}}=\eta_{c}|h_{c_{k}d_{k}}|^{2}$.

We hereby employ an opportunistic IoT transmitter-receiver pair selection ($C_{k^{\ast}}$$-$$D_{k^{\ast}}$) strategy that maximizes the SINR at satellite receiver $B$ i.e.,
$k^{\ast}=\arg \displaystyle \max_{k\in\{1,...,K\}} \Lambda_{ac_{k}b}$.

\section{Outage Probability of Satellite Network}\label{per}
In this section, we evaluate the OP and achievable diversity order of the satellite network of considered OSTN.

For a target rate $\mathcal{R}_{p}$, the outage probability of primary satellite network with selected IoT network $k^{\ast}$ is obtained as
\begin{align}\label{cmqr}
\mathcal{P}^{\textmd{sat}}_{\textmd{out}}(\mathcal{R}_{p}) 
&=\textmd{Pr}\left[\Lambda_{ac_{k^{\ast}}b}<\gamma_{p}\right]\\\nonumber
&=\mathbb{E}\{\left(\textmd{Pr}\left[\Lambda_{ac_{k}b}<\gamma_{p}|W_c=w\right]\right)^K\},
\end{align}
where $\gamma_{p}=2^{2\mathcal{R}_{p}}-1$ and $\mathbb{E}\{\cdot\}$ is the statistical expectation. Since the foregoing analysis is intractable based on the exact SINR in (\ref{snrs}), we derive a tight lower bound on the exact OP of satellite network.    
\newtheorem{theorem}{Theorem}
\begin{theorem}\label{tth2}
	The tight lower bound outage probability of satellite network $\tilde{\mathcal{P}}^{\textmd{sat}}_{\textmd{out}}(\mathcal{R}_{p})$ can be given as
	\begin{align}\label{asl23}
	\tilde{\mathcal{P}}^{\textmd{sat}}_{\textmd{out}}(\mathcal{R}_{p})&=\left\{ \begin{array}{l}
	\Psi(\mathcal{R}_{p}),\textmd{ if } \gamma_{p} < \mu^{\prime} \\
	1, \textmd{ if } \gamma_{p} \geq\mu^{\prime}
	\end{array}\right.,
	\end{align}
	where $\Psi(\mathcal{R}_{p})$ is given as 
		\begin{align}\label{si3}
		&\Psi(\mathcal{R}_{p})=\sum_{n=0}^{K}\binom{K}{n}(-1)^n\alpha^n_c\sum_{s_m\in\mathcal{S}}\frac{n!}{\prod_{m=0}^{m_{ac}-1}s_m!}\\\nonumber
		&\times\prod_{m=0}^{m_{ac}-1}(\mathcal{A}_m)^{s_m}
		\tilde{\gamma}^{\Delta_{ac}}_p\textmd{e}^{-\frac{n\tilde{\gamma}_p}{\Omega_{cb}\eta_c}}\widetilde{\sum}\frac{\Xi(M_1)}{\eta^{\Lambda}_s}\left(\frac{1}{\Omega_t\eta_t}\right)^{M_2}\\\nonumber
		&\times\frac{\Phi(M_2,\Lambda)}{\Gamma(M_2)}\frac{\Gamma(M_2+\Lambda)}{\Gamma(\Lambda)}\left(n\tilde{\gamma_p}\Theta_{c,a}+\frac{1}{\Omega_t\eta_t}\right)^{-(\Delta_{ac}+\Lambda+M_2)}\\\nonumber
		&\times\MeijerG{1}{2}{2}{2}{\frac{\Theta_{s,s}}{n\tilde{\gamma_p}\Theta_{c,a}+\frac{1}{\Omega_t \eta_t}}}{1-\Delta_{ac}-\Lambda-M_2, 1-\Lambda}{0,1-\Lambda-M_2},
		\end{align}
with $\mu^{\prime}=\frac{\mu}{1-\mu}$, $\tilde{\gamma}_{p}=\frac{\gamma_{p}}{\mu-(1-\mu)\gamma_{p}}$,  $\mathcal{S}=\{S_{m}|{\sum}_{m=0}^{m_{ac}-1}s_{m}=n\}$ , $\Delta_{ac}={\sum}_{m=0}^{m_{ac}-1}m s_{m}$, $\mathcal{A}_{m}={\sum}_{l=m}^{m_{ac}-1}\frac{\zeta(l)}{(\eta_{a})^{l+1}}\frac{l!}{m!} (\Theta_{c,a})^{-(l+1-m)}$, $\{s_{m}\}$ are nonnegative integers, $\Theta_{s,s}=\tilde{\Theta}_{s,s}-\frac{1}{\Omega_t\eta_t}$, $\tilde{\Theta}_{s,s}=\frac{\beta_s-\delta_s}{\eta_s}$, $\Theta_{c,a}=\frac{\beta_c-\delta_c}{\eta_a}$, $\widetilde{\sum}=\sum_{i_1}^{m_s-1}\dots\sum_{i_{M_1}}^{m_s-1}$, and  $\Phi(\cdot,\cdot)$ and $\Gamma(\cdot)$ denote the Beta function \cite[eq. 8.384.1]{grad} and gamma function \cite[eq. 8.350.1]{grad}, respectively along with $G^{1,2}_{2,2}[\cdot]$ as univariate Meijer's-$G$ function \cite[eq. 8.2.1.1]{grad}.
	\end{theorem}
\begin{IEEEproof}
	See Appendix \ref{appA}.
\end{IEEEproof}
Note that, in (\ref{asl23}), $\gamma_{p} < \mu^{\prime}$ is the necessary condition to allow secondary spectrum access for IoT network, otherwise an outage event is induced.

In the following corollary, we derive the asymptotic OP for the IoT-assisted satellite network to reveal its diversity order.
\newtheorem{corollary}{Corollary}
\begin{corollary}\label{cor1}
	The asymptotic OP for satellite network under $\gamma_{p} < \mu^{\prime}$ and $\eta_a=\eta_c=\eta$ can be given as
	\begin{align}\label{asyp}
	\tilde{\mathcal{P}}^{\textmd{sat}}_{\textmd{out},\infty}(\mathcal{R}_{p})&=
		\sum_{n=0}^{K}\binom{K}{n}\Big(\frac{\alpha^n_c}{\Omega^{K-n}_{cb}}\Big)\Big(\frac{\tilde{\gamma}_p}{\eta}\Big)^K\psi(n),
	\end{align}
	where the function $\psi(n)$ is defined as
	\begin{align}
	&\psi(n)=\widetilde{\sum}\frac{\Xi(M_1)}{\eta^{\Lambda}_s}\Big(\frac{1}{\Omega_t\eta_t}\Big)^{-(n+\Lambda)}\frac{\Phi(M_2,\Lambda)}{\Gamma(M_2)}\\\nonumber
	&\times\frac{\Gamma{(M_2+\Lambda)}}{\Gamma(\Lambda)}\MeijerG{1}{2}{2}{2}{{\Theta_{s,s}\Omega_t\eta_t}}{1-\Lambda-M_2-n, 1-\Lambda}{0,1-\Lambda-M_2}.
	\end{align}
\end{corollary}
\begin{IEEEproof}
The proof follows the similar steps as given in Appendix \ref{appA}. Under $\eta\rightarrow\infty$, we can approximate  (\ref{hsi}) as
\begin{align}\label{ahsi}
\tilde{\mathcal{P}}^{\textmd{sat}}_{\textmd{out}}(\mathcal{R}_{p})&=\mathbb{E}\{(F_{\hat{\Lambda}_{ac_k}}(\tilde{\gamma}_p|W_c=w)+F_{\Lambda_{c_kb}}(\tilde{\gamma}_p))^K\},
\end{align}
where the product term leading to the higher order is neglected. Further, at $\eta\rightarrow\infty$, we can simplify the cdfs $F_{\hat{\Lambda}_{ac_{k}}}(x|W_c=w)\simeq \frac{\alpha_{c}xw}{\eta}$ and $F_{\Lambda_{c_{k}b}}(x)\simeq\big(\frac{x}{\Omega_{cb}\eta}\big)$, for small argument $x$. Subsequently, substituting these cdfs into (\ref{ahsi}) results in (after applying the binomial expansion) the following expression
\begin{align}
	&\tilde{\mathcal{P}}^{\textmd{sat}}_{\textmd{out},\infty}(\mathcal{R}_{p})=\sum_{n=0}^{K}\binom{K}{n}\!\Big(\frac{\alpha^n_c}{\Omega^{K-n}_{cb}}\Big)\!\Big(\frac{\tilde{\gamma}_p}{\eta}\Big)^{K}\\\nonumber
	&\times\widetilde{\sum}\frac{\Xi(M_1)}{\eta^{\Lambda}_s}\Big(\frac{1}{\Omega_t\eta_t}\Big)^{M_2}\frac{\Phi(M_2,\Lambda)}{\Gamma(M_2)}\frac{\Gamma{(M_2+\Lambda)}}{\Gamma(\Lambda)}\\\nonumber
	&\times \int_{0}^{\infty}w^{n+\Lambda+M_2-1}\textmd{e}^{-\frac{w}{\Omega_t\eta_t}}\MeijerG{1}{1}{1}{2}{\Theta_{s,s}w}{1-\Lambda}{0,1-\Lambda-M_2}dw,
\end{align}
where the Meijer-G representation of function ${_1F}_1(\cdot;\cdot;\cdot)$ has also been applied \cite[eq. 07.20.26.0006.01]{wolf}. Finally, on evaluating the resulting integral with the aid of \cite[eq. 7.813]{grad}, we can achieve (\ref{asyp}).  
\end{IEEEproof}

\emph{Remark 1:} Upon re-expressing (\ref{asyp}) as $\mathcal{G}_c\eta^{-\mathcal{G}_d}$ with $\eta_s=\eta_t$ set fixed, and neglecting the higher order terms, the achievable diversity order $\mathcal{G}_d$ of satellite network is $K$. However, if interferers' power varies proportional to $\eta$, i.e., $\eta_s=\eta_t=\nu\eta$ for some constant $\nu$, $\mathcal{G}_d$ reduces to zero.   

\section{Outage Probability of IoT Network}\label{sec}
In this section, we evaluate the OP and achievable diversity order of the IoT network of considered OSTN.

Given a target rate $\mathcal{R}_{\textmd{S}}$, based on the SINR in (\ref{secb}), the OP of secondary IoT network can be computed as
\begin{align}\label{cm0sec}
&\mathcal{P}^\textmd{IoT}_{\textmd{out}}(\mathcal{R}_{\textmd{S}}) =\textmd{Pr}[{\Lambda}_{a{c_{k^\ast}}{d_{k^\ast}}}<\gamma_s]\\\nonumber
&=\mathbb{E}\Big\{\textmd{Pr}\bigg[\frac{\mu\hat{\Lambda}_{c_{k^{\ast}}d_{k^{\ast}}}(\hat{\Lambda}_{ac_{k^{\ast}}}+1)}{\mu\hat{\Lambda}_{c_{k^{\ast}}d_{k^{\ast}}}+\hat{\Lambda}_{ac_{k^{\ast}}}+1}
<\mu^{\prime}\gamma_{s}\Big|W_c=w\bigg]\Big\},
\end{align}
where $\gamma_{s}=2^{2\mathcal{R}_{\textmd{S}}}-1$. Hereby, for a tractable analysis, we apply the bound $\frac{XY}{X+Y}\leq\min(X,Y)$ to evaluate the lower bound on OP $\tilde{\mathcal{P}}^\textmd{IoT}_{\textmd{out}}(\mathcal{R}_{\textmd{S}})$ in (\ref{cm0sec}) as
\begin{align}\label{cm0sec1}
\tilde{\mathcal{P}}^\textmd{IoT}_{\textmd{out}}(\mathcal{R}_{\textmd{S}}) &=\mathbb{E}\{\tilde{\mathcal{P}}^\textmd{IoT}_{\textmd{out}}(\mathcal{R}_{\textmd{S}}|W_c=w)\},
\end{align}
where the conditional OP $\tilde{\mathcal{P}}^\textmd{IoT}_{\textmd{out}}(\mathcal{R}_{\textmd{S}}|w)$ can be expressed as 
\begin{align}
\tilde{\mathcal{P}}^\textmd{IoT}_{\textmd{out}}(\mathcal{R}_{\textmd{S}}|w)&=\textmd{Pr}[\min{(\mu\hat{\Lambda}_{c_{k^{\ast}}d_{k^{\ast}}},\hat{\Lambda}_{ac_{k^{\ast}}}\!\!+\!1)}
\!<{\mu^{\prime}\gamma_{s}}|w].
\end{align}
After a simple variable transformation for random variable $\hat{\Lambda}_{ac_{k^{\ast}}}+1$, the OP in (\ref{cm0sec1}) can be further expressed as
\begin{align}\label{aql23}
\tilde{\mathcal{P}}^\textmd{IoT}_{\textmd{out}}(\mathcal{R}_{\textmd{S}}|w)&=\left\{ \begin{array}{l}
F_{\mu\hat{\Lambda}_{c_{k^{\ast}}d_{k^{\ast}}}}(\mu^{\prime}\gamma_{s}|w),\textmd{ if } \gamma_{s} < \frac{1}{\mu^{\prime}} \\
1-\overline{F}_{\mu\hat{\Lambda}_{c_{k^{\ast}}d_{k^{\ast}}}}(\mu^{\prime}\gamma_{s}|w)
\\\times\overline{F}_{\hat{\Lambda}_{ac_{k^{\ast}}}}({\mu^{\prime}\gamma_{s}}-1|w), \textmd{ if } \gamma_{s} \geq \frac{1}{\mu^{\prime}}
\end{array}\right.
\end{align}
where $\overline{F}_{X}(\cdot|w)=1-F_{X}(\cdot|w)$. We evaluate the OP given by (\ref{cm0sec1}) in the following theorem.

\newtheorem{lemma}{Lemma}
\begin{theorem}\label{th1}
	The lower bound on OP of secondary IoT network $\tilde{\mathcal{P}}^{\textmd{IoT}}_{\textmd{out}}(\mathcal{R}_{s})$ is given as
\begin{align}\label{ajsl23}
\tilde{\mathcal{P}}^{\textmd{IoT}}_{\textmd{out}}(\mathcal{R}_{s})&=\left\{ \begin{array}{l}
\Psi_1(\mathcal{R}_{s}),\textmd{ if } \gamma_{s} < \frac{1}{\mu^{\prime}}, \\
\Psi_1(\mathcal{R}_{s})+\Psi_2(\mathcal{R}_{s})+\Psi_3(\mathcal{R}_{s}),\\\textmd{ if } \gamma_{s} \geq\frac{1}{\mu^{\prime}},
\end{array}\right.
\end{align}
where $\Psi_1(\mathcal{R}_{s})$ and $\Psi_2(\mathcal{R}_{s})$ are given as (\ref{si11}) and (\ref{si22}), respectively, on the next page. 
\begin{figure*}[!t]
	\begin{align}\label{si11}
	\Psi_1(\mathcal{R}_{s})&=1-\widetilde{\sum}\frac{\Xi(M_1)}{\eta^{\Lambda}_s}\left(\frac{1}{\Omega_t\eta_t}\right)^{M_2}\frac{\Phi(M_2,\Lambda)}{\Gamma(M_2)}\frac{\Gamma(M_2+\Lambda)}{\Gamma(\Lambda)}\chi^{-(\Lambda+M_2)}_{c,t}\MeijerG{1}{2}{2}{2}{\frac{\Theta_{s,s}}{\chi_{c,t}}}{1-\Lambda-M_2, 1-\Lambda}{0,1-\Lambda-M_2}.
	\end{align}
	\hrule
\end{figure*}
\begin{figure*}[!t]
	\begin{align}\label{si22}
	\Psi_2(\mathcal{R}_{s})&=K\sum_{l=0}^{m_{ac}-1}\frac{\zeta(l)}{\eta^{l+1}_a}\sum_{n=0}^{K-1}\binom{K-1}{n}(-1)^n\alpha^{n+1}_c\sum_{s_m\in\mathcal{S}}\frac{n!}{\prod_{m=0}^{m_{ac}-1}s_m!}\prod_{m=0}^{m_{ac}-1}(\mathcal{A}_m)^{s_m}\widetilde{\sum}\frac{\Xi(M_1)}{\eta^{\Lambda}_s}\left(\frac{1}{\Omega_t\eta_t}\right)^{M_2}\\\nonumber
	&\times\frac{\Phi(M_2,\Lambda)}{\Gamma(M_2)}\frac{\Gamma(M_2+\Lambda)}{\Gamma(\Lambda)}\left({\frac{n+1}{\Omega_{cb}\eta_c}}\right)^{-(\tilde{l}+1)}\left[\chi^{-\tilde{\Delta}_{ac}}_{c,t}\MeijerGG{1,1,1,1,1}{1,[1:1],0,[1:2]}{\frac{\Theta_{c,a}\Omega_{cb}\eta_c}{\chi_{c,t}}}{\frac{\Theta_{s,s}}{\chi_{c,t}}}
	{\tilde{\Delta}_{ac};-\tilde{l};1-\Lambda}{-;0;0;1-\Lambda-M_2}\right.\\\nonumber
	&-\sum_{q=0}^{\tilde{l}}\frac{\tilde{\gamma}^q_s}{q!}\textmd{e}^{-\frac{(n+1)\tilde{\gamma}_s}{\Omega_{cb}\eta_c}}\left(\frac{n+1}{\Omega_{cb}\eta_c}\right)^{q}\frac{\Gamma(\tilde{l}+1)}{\Gamma(\tilde{l}+1-q)}\tilde{\chi}^{-\tilde{\Delta}_{ac}}_{c,t}\left.\MeijerGG{1,1,1,1,1}{1,[1:1],0,[1:2]}{\frac{\Theta_{c,a}\Omega_{cb}\eta_c}{\tilde{\chi}_{c,t}}}{\frac{\Theta_{s,s}}{\tilde{\chi}_{c,t}}}
	{\tilde{\Delta}_{ac};-\tilde{l}+q;1-\Lambda}{-;0;0;1-\Lambda-M_2}\right].
	\end{align}
	\hrule
\end{figure*}
Further, $\Psi_3(\mathcal{R}_{s})$ is given by
\begin{align}\label{si33}
&\Psi_3(\mathcal{R}_{s})=\frac{K}{\Omega_{cb}\eta_c}\sum_{l=0}^{m_{ac}-1}\frac{\zeta(l)}{\eta^{l+1}_a}\sum_{n=0}^{K-1}\binom{K-1}{n}(-1)^n\\\nonumber
&\times\alpha^{n+1}_c\sum_{s_m\in\mathcal{S}}\frac{n!}{\prod_{m=0}^{m_{ac}-1}s_m!}\prod_{m=0}^{m_{ac}-1}(\mathcal{A}_m)^{s_m}\widetilde{\sum}\frac{\Xi(M_1)}{\eta^{\Lambda}_s}\\\nonumber
&\times\left(\frac{1}{\Omega_t\eta_t}\right)^{M_2}\frac{\Phi(M_2,\Lambda)}{\Gamma(M_2)}\Gamma(\Delta_{ac}+1)\left(\Psi_4(\mathcal{R}_{s})-\Psi_5(\mathcal{R}_{s})\right),
\end{align}
where $\Psi_4(\mathcal{R}_{s})$ and $\Psi_5(\mathcal{R}_{s})$ are given by (\ref{si44}) and (\ref{si55}), respectively, at the top of next page. In Theorem \ref{th1}, the various terms are denoted as $\tilde{\gamma}_{s}={\mu^{\prime}\gamma_{s}-1}$, $\chi_{c,t}=\frac{1}{\Omega_t\eta_t}+\frac{\gamma_s}{\Omega_{cd}\eta_c(1-\mu)}$, $\chi_{c,s}=\frac{\beta_s-\delta_s}{\eta_s}+\frac{\gamma_s}{\Omega_{cd}\eta_c(1-\mu)}$, $\tilde{\Lambda}=\Delta_{ac}+\Lambda+M_2$, $\tilde{\Delta}_{ac}=\tilde{\Lambda}+l+1$, $\tilde{l}=l+\Delta_{ac}$, $\tilde{g}=\tilde{\Delta}_{ac}-\Lambda-g$, $\tilde{\chi}_{c,t}={\left(\chi_{c,t}+(n+1)\Theta_{c,a}\tilde{\gamma}_s\right)}$, $\tilde{\chi}_{c,s}=\left(\chi_{c,s}+(n+1)\Theta_{c,a}\tilde{\gamma}_s\right)$, with $G^{1,1,1,1,1}_{1,[1:1],0,[1:1]}[\cdot]$ and $G^{1,1,1,1,1}_{1,[1:1],0,[1:2]}[\cdot]$ as the bivariate Meijer's $G$-functions  \cite{kangc}, respectively. The rest of the terms are the same as defined previously in Theorem \ref{tth2}.
	\begin{figure*}[!t]
	\begin{align}\label{si44}
	\Psi_4(\mathcal{R}_{s})&=\Theta^{-(l+1)}_{c,a}\frac{\Gamma(M_2+\Lambda){\Gamma(l+1)}}{\Gamma(\Lambda)\Gamma(\Delta_{ac}+1)}\left(\frac{n\!+\!1}{\Omega_{cb}\eta_c}\right)^{-(\Delta_{ac}+1)}\left[\chi^{-\tilde{\Lambda}}_{c,t}\MeijerGG{1,1,1,1,1}{1,[1:1],0,[1:2]}{\frac{n\Theta_{c,a}\Omega_{cb}\eta_c}{(n+1){\chi}_{c,t}}}{\frac{\Theta_{s,s}}{{\chi}_{c,t}}}
	{\tilde{\Lambda};-\Delta_{ac};1-\Lambda}{-;0;0;1-\Lambda-M_2}\right.\\\nonumber
	&-\sum_{u=0}^{l}\frac{(\Theta_{c,a}\tilde{\gamma}_s)^u}{u!}\left(\chi_{c,t}+\Theta_{c,a}\tilde{\gamma}_s\right)^{-(\tilde{\Lambda}+u)}\left.\MeijerGG{1,1,1,1,1}{1,[1:1],0,[1:2]}{\frac{n\Theta_{c,a}\Omega_{cb}\eta_c}{(n+1)(\chi_{c,t}+\Theta_{c,a}\tilde{\gamma}_s)}}{\frac{\Theta_{s,s}}{\chi_{c,t}+\Theta_{c,a}\tilde{\gamma}_s}}
	{\tilde{\Lambda}+u;-\Delta_{ac};1-\Lambda}{-;0;0;1-\Lambda-M_2}\right].
	\end{align}
	\hrule
\end{figure*}
	\begin{figure*}[!t]
	\begin{align}\label{si55}
	\Psi_5(\mathcal{R}_{s})&=\sum_{q=0}^{\Delta_{ac}}\sum_{g=0}^{M_2}\binom{M_2-1}{g}\frac{(-1)^{-(2\Lambda+g)}\Theta^{-(\Lambda+g)}_{s,s}\Gamma(\Lambda+g)}{\Phi(\Lambda, M_2)q!\Gamma(\Delta_{ac}-q+1)}\left[\chi^{-\tilde{g}}_{c,t}\MeijerGG{1,1,1,1,1}{1,[1:1],0,[1:1]}{{\frac{n\Theta_{c,a}\Omega_{cb}\eta_c}{(n+1)\chi_{c,t}}}}{{\frac{\Theta_{c,a}\Omega_{cb}\eta_c}{\chi_{c,t}}}}
	{\tilde{g};-{\Delta}_{ac}+{q};-l-q}{-;0;0}\right.\\\nonumber
	&-\sum_{j=1}^{l+q}\frac{\tilde{\gamma}^{j}_{s}}{j!}\textmd{e}^{-\frac{(n+1)\tilde{\gamma}_s}{\Omega_{cb}\eta_c}}\left(\frac{n+1}{\Omega_{cb}\eta_c}\right)^{j}
	\tilde{\chi}^{-\tilde{g}}_{c,t}\frac{\Gamma(l+q+1)}{\Gamma(l+q-j+1)}\MeijerGG{1,1,1,1,1}{1,[1:1],0,[1:1]}{{\frac{n\Theta_{c,a}\Omega_{cb}\eta_c}{(n+1)\tilde{\chi}_{c,t}}}}{{\frac{\Theta_{c,a}\Omega_{cb}\eta_c}{\tilde{\chi}_{c,t}}}}
	{\tilde{g};-{\Delta}_{ac}+q;-l-q+j}{-;0;0}\\\nonumber
	&-\sum_{v=0}^{\Lambda+g-1}\frac{\Theta^{v}_{s,s}}{v!}\left[\chi^{-(\tilde{g}+v)}_{c,s}\MeijerGG{1,1,1,1,1}{1,[1:1],0,[1:1]}{{\frac{n\Theta_{c,a}\Omega_{cb}\eta_c}{(n+1)\chi_{c,s}}}}{{\frac{\Theta_{c,a}\Omega_{cb}\eta_c}{\chi_{c,s}}}}
	{\tilde{g}+v;-{\Delta}_{ac}+q;-l-q}{-;0;0}\right.-\sum_{j=1}^{l+q}\frac{\tilde{\gamma}^{j}_{s}}{j!}\textmd{e}^{-\frac{(n+1)\tilde{\gamma}_s}{\Omega_{cb}\eta_c}}\left(\frac{n+1}{\Omega_{cb}\eta_c}\right)^{j}\\\nonumber
	&\times\frac{\Gamma(l+q+1)}{\Gamma(l+q-j+1)}\tilde{\chi}^{-(\tilde{g}+v)}_{c,s}\left.\left.\MeijerGG{1,1,1,1,1}{1,[1:1],0,[1:1]}{{\frac{n\Theta_{c,a}\Omega_{cb}\eta_c}{(n+1)\tilde{\chi}_{c,s}}}}{{\frac{\Theta_{c,a}\Omega_{cb}\eta_c}{\tilde{\chi}_{c,s}}}}
	{\tilde{g}+v;-{\Delta}_{ac}+q;-l-q+j}{-;0;0}\right]\right]\left(\frac{n+1}{\Omega_{cb}\eta_c}\right)^{-(\tilde{l}+2)}.
 \end{align}
	\hrule
\end{figure*}
\end{theorem}
\begin{IEEEproof}
In (\ref{aql23}), the cdf $F_{\mu\hat{\Lambda}_{c_{k^{\ast}}d_{k^{\ast}}}}(x|w)$ can be written as $F_{\mu\hat{\Lambda}_{c_{k}d_{k}}}(x|w)=1-\textmd{e}^{-\frac{xw}{\Omega_{cd}\eta_c\mu}}$ (for $W_c+1\approx W_c$ under high interference power level) since the selection strategy is independent of the $C_k-D_k$ link. However, the cdf ${F}_{\hat{\Lambda}_{ac_{k^{\ast}}}}(x|w)$ follows the order statistics according to the considered selection strategy, which can be calculated as
\begin{align}\label{apjkera}
{F}_{\hat{\Lambda}_{ac_{k^{\ast}}}}\!(x|w)\!&=\! K\textmd{Pr}\Big[\hat{\Lambda}_{ac_{k}}\!\!<\!x,{\Lambda}_{ac_{k}b}\!>\!\!\! \displaystyle\max_{\substack { l=1,...,K\\ l \neq k}}{\Lambda}_{ac_{l}b}|w\Big].
\end{align}
As such, (\ref{apjkera}) is tedious to solve exactly. So, we first simplify the SINR ${\Lambda}_{ac_kb}$ as in Appendix \ref{appA}, and then, follow the steps as given in \cite[Appendix C]{pkss}, to get the cdf ${F}_{\hat{\Lambda}_{ac_{k^{\ast}}}}(x|w)$ as (\ref{stat}) on the next page, where $\vartheta_n=n\Theta_{c,a}w+\frac{n+1}{\Omega_{cb}\eta_c}$, $\omega_n=(n+1)(\Theta_{c,a}w+\frac{1}{\Omega_{cb}\eta_c})$, and $\Upsilon(\cdot,\cdot)$ is the lower incomplete gamma function \cite[eq. 8.310.1]{grad}. 
Finally, evaluating (\ref{aql23}) with the help of (\ref{stat}), and utilizing the result in (\ref{cm0sec1}) (with some involved mathematical manipulations, e.g., series representation of lower incomplete gamma function \cite[eq. 8.352.6]{grad} and ${_1F}_1(\cdot;\cdot;\cdot)$ \cite[eq. 07.20.03.0106.01]{wolf} along with the Meijer-G representation of ${_1F}_1(\cdot;\cdot;\cdot)$ \cite[eq. 07.20.26.0006.01]{wolf}), we get the closed-form OP of IoT network using \cite[eq. 07.34.21.0081.01]{wolf}. 
\begin{figure*}[!t]
	\begin{align}\label{stat}
	F_{\hat{\Lambda}_{ac_{k^{\ast}}}}(x|w)&=\frac{K}{\Omega_{cb}\eta_c}\sum_{n=0}^{\mathrm{K}-1}
	\binom{K-1}{n}(-1)^{n}\alpha^{n+1}_{c}\sum_{S_{m}\in \mathcal{S}}\frac{n!}{{\prod}_{m=0}^{m_{ac}-1}s_{m}!}
	\prod_{m=0}^{m_{ac}-1}\left(\mathcal{A}_{m}\right)^{s_{m}}\frac{\Gamma(\Delta_{ac}+1)}
	{\vartheta^{\Delta_{ac}+1}_n}
	\sum_{l=0}^{m_{ac}-1}\frac{\zeta(l)}{\eta^{l+1}_{a}} \\\nonumber
	&\times w^{\Delta_{ac}+l+1}	
	\Bigg[\frac{\Upsilon(l+1,\Theta_{c,a}wx)}
	{(\Theta_{c,a}w)^{l+1}}-\sum_{q=0}^{\Delta_{ac}}
	\frac{\left(\vartheta_{n}\right)^{q}}{q!\omega^{l+q+1}_{n}}
	{\Upsilon\left(l+q+1,\omega_{n}x\right)}
	\Bigg]+\mathrm{K}\sum_{l=0}^{m_{ac}-1}\frac{\zeta(l)}{\eta^{l+1}_{a}}\\\nonumber
	&\times
	\sum_{n=0}^{\mathrm{K}-1}
	\binom{K-1}{n}(-1)^{n}\alpha^{n+1}_{c}\sum_{S_{m}\in \mathcal{S}}\frac{n!}{{\prod}_{m=0}^{m_{ac}-1}s_{m}!}\prod_{m=0}^{m_{ac}-1}\left(\mathcal{A}_{m}\right)^{s_{m}}
	w^{\Delta_{ac}+l+1}
	\frac{\Upsilon\left(l+\Delta_{ac}+1,\omega_{n}x\right)}
	{\omega^{l+\Delta_{ac}+1}_{n}}.
	\end{align}
	\hrule
\end{figure*}	
\end{IEEEproof}

Next, we examine the asymptotic OP of IoT network for achievable diversity order.
\begin{corollary}\label{cor2}
	The asymptotic OP for secondary IoT network under $\gamma_{p} < \mu^{\prime}$ and $\eta_a=\eta_c=\eta$ can be given as
\begin{align}\label{asyy2}
\tilde{\mathcal{P}}^{\textmd{IoT}}_{\textmd{out},\infty}(\mathcal{R}_{s})&\simeq\left\{ \begin{array}{l}
\frac{1}{\Omega_{cd}\eta}(\frac{\gamma_{s}}{1-\mu})\psi(1), \textmd{ if } \gamma_{s} < \frac{1}{\mu^{\prime}},\\
\sum_{n=0}^{K-1}\binom{K-1}{n}(\frac{\alpha^{n+1}_c}{\Omega^{K\!-\!n\!-\!1}_{cb}})(\!\frac{\tilde{\gamma}_s}{\eta}\!)^K\psi(\!n\!\!+\!\!1\!)\\
+\frac{1}{\Omega_{cd}\eta}(\frac{\gamma_{s}}{1-\mu})\psi(1), \textmd{ if } \gamma_{s} \geq \frac{1}{\mu^{\prime}},
\end{array}\right.
\end{align}
where the function $\psi(\cdot)$ is already defined in Corollary \ref{cor1}. 
\end{corollary}
\begin{IEEEproof}
To derive asymptotic OP of IoT network according to (\ref{cm0sec1}), we need to express $\tilde{\mathcal{P}}^\textmd{IoT}_{\textmd{out}}(\mathcal{R}_{\textmd{S}}|w)$ in (\ref{aql23}). For this, we can simplify the cdf $F_{\mu\hat{\Lambda}_{c_{k^{\ast}}d_{k^{\ast}}}}(x|w) =(\frac{xw}{\Omega_{cd}\eta\mu})$ and the cdf $F_{\hat{\Lambda}_{ac_{k^{\ast}}}}(x|w)$ for small $x$ (based on (\ref{apjkera})) as
\begin{align}\label{stgh}
F_{\hat{\Lambda}_{ac_{k^{\ast}}}}(x|w)&\simeq
\Big(\frac{\alpha_{c}w}{\eta}\Big)\Big(\frac{\alpha_{c}w}{\eta}\!+\!\frac{1}{\Omega_{cb}\eta}\Big)^{K\!-\!1}x^{K}. \end{align}
Now, first, invoking these cdfs in (\ref{aql23}), and then, substituting the result in (\ref{cm0sec1}), we can reach at (\ref{asyy2}) after evaluating the underlying expectation similar to that in Corollary \ref{cor1}.  	
\end{IEEEproof}

\emph{Remark 2:} Upon re-expressing (\ref{asyy2}) as $\mathcal{G}_c\eta^{-\mathcal{G}_d}$ with $\eta_s=\eta_t$ fixed, and neglecting the higher order terms, one can observe that the achievable diversity order $\mathcal{G}_d$ of IoT network is unity. Moreover, similar to the case of satellite network, if interferers' power varies proportional to $\eta$, i.e., $\eta_s=\eta_t=\nu\eta$ for some constant $\nu$, $\mathcal{G}_d$ reduces to zero. Hereby, the impact of $K$ on diversity order of IoT network is not reflected.
\section{Adaptive Power Splitting Factor}\label{psf}
In this section, we devise the scheme for finding the appropriate value of power splitting factor $\mu$ for effective spectrum sharing. Recalling the necessary condition $\gamma_{p} < \mu^{\prime}$ in Theorem \ref{tth2}, the feasible dynamic range of $\mu$ can be formulated as $\frac{\gamma_{\textmd{p}}}{1+\gamma_{\textmd{p}}} \leq \mu \leq 1$. Further, to obtain $\mu$, a quality-of-service (QoS) constraint must be imposed to protect the satellite network from IoT transmissions. Thus, we choose the value of $\mu$ such that the OP of satellite network $\tilde{\mathcal{P}}^{\textmd{sat}}_{\textmd{out}}(\mathcal{R}_{p})$ is guaranteed below a predetermined QoS level $\epsilon$, i.e., $\tilde{\mathcal{P}}^{\textmd{sat}}_{\textmd{out}}(\mathcal{R}_{p})\leq \epsilon$. Note that if this QoS constraint is taken at equality, the resulting value of $\mu$ minimizes the OP of IoT network (i.e., $\tilde{\mathcal{P}}^{\textmd{IoT}}_{\textmd{out}}(\mathcal{R}_{s})$). Although the closed-form solution of $\mu$ under above constraints is infeasible, it can be determined via numerical search method. Moreover, we consider the case of assigning an arbitrary fixed value of $\mu$ within its dynamic range for comparison. 
\section{Numerical and Simulation Results}\label{num}
In this section, we present numerical results for considered OSTN. We further validate the theoretical results via simulations for $10^6$ independent channel realizations. We set $\mathcal{R}_{p}=\mathcal{R}_{\textmd{S}}=0.5$ bps/Hz so that $\gamma_{p}=1$, $\gamma_{s}=1$ (unless stated otherwise), $M_1=M_2=2$, $\Omega_{cb}=\Omega_{cd}=1$, and $\eta_{a}=\eta_{c}=\eta$ as SNR. The shadowed-Rician fading parameters for satellite link $A-C_{k}$ and extra terrestrial interferers are considered as ($m_{ac},\flat_{ac},\Omega_{ac}=5,0.251,0.279$) (for light shadowing) and ($m_{s},\flat_{s},\Omega_{s}=2,0.063,0.0005$) (for heavy shadowing), respectively. We further set $\Omega_t=0.2$ for terrestrial interferers, and interferers' power $\eta_s=\eta_t=20$ dB. Here, we consider two cases (a) when interferers' power is fixed, i.e., $\eta_s=\eta_t=20$ dB, and (b) when interferers' power is proportional to $\eta$, i.e.,  $\eta_s=\eta_t=\nu\eta$, with $\nu=-15$ dB.

\begin{figure}
\centering
\includegraphics[width=3.3in]{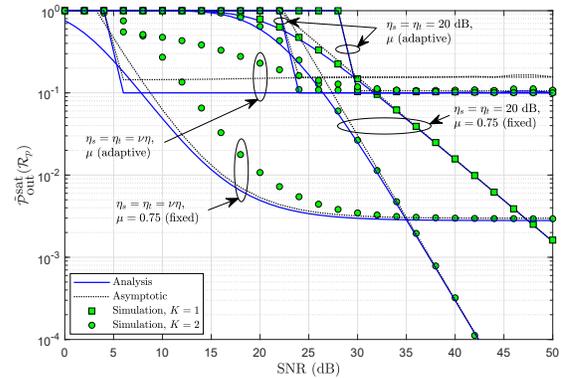}
\caption{OP of satellite network versus SNR.}
\label{fig1}
\end{figure}

In Fig.~\ref{fig1}, we plot the OP curves for the satellite network for various system parameters. We plot the set of curves corresponding to fixed value of power splitting factor (i.e., $\mu=0.75$) as well as its adaptive values calculated according to section \ref{psf}. For fixed $\eta_s$, $\eta_t$ and $\mu$, the analytical lower bound OP curves are very tight to the simulation results. Further, the diversity order of $K$ is confirmed for the satellite network by the slope of corresponding asymptotic OP curves as $K$ changes form $1$ to $2$. In contrast, if $\eta_s$, $\eta_t$ varies proportional to $\eta$, the diversity order of the satellite network reduces to zero (see flat OP curve, $K=2$). Hereby, the analytical OP curves are tight in medium to SNR region only. Furthermore, if adaptive procedure for $\mu$ is followed, the OP of satellite network can be guaranteed to a predetermined QoS level ($\epsilon=0.1$) beyond a certain value of SNR (e.g., $24$ dB and $29$ dB for $K=2$ and $1$, respectively), otherwise the outage event occurs. Nonetheless, the diversity gain may not be harvested with adaptive $\mu$ since the QoS level $\epsilon$ is fixed to  $0.1$.  
    
\begin{figure}
	\centering
	\includegraphics[width=3.3in]{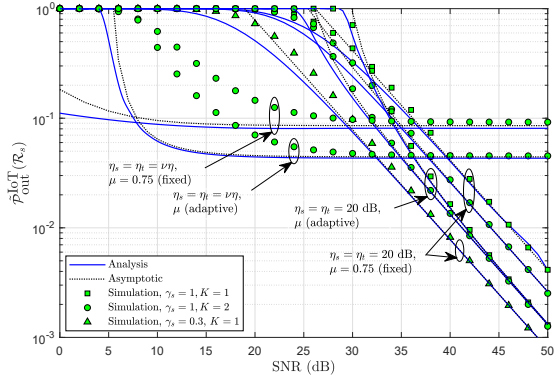}
	\caption{OP of IoT network versus SNR.}
	\label{fig2}
\end{figure}

In Fig.~\ref{fig2}, we plot the OP curves for the IoT network for various system parameters. We set the value of $\gamma_s=1$ and $0.3$ to obtain $\gamma_s\geq\frac{1}{\mu^\prime}$ and $\gamma_s<\frac{1}{\mu^\prime}$ according to Theorem \ref{th1}. Clearly, for fixed values of $\eta_s$, $\eta_t$ and $\mu$, we can observe that the lower bound analytical OP curves are in close proximity to simulation results. Note that only the unity diversity order is achievable for IoT network as seen by the same slope of asymptotic OP curves for different $K$ (i.e., $1$ and $2$). More importantly, if $\mu$ is chosen adaptively, the OP of IoT network improves remarkably beyond a certain value of SNR (e.g., $24$ dB and $29$ dB for $K=2$ and $1$, respectively) without affecting the achievable diversity order (for fixed $\eta_s$ and $\eta_t$), and prior to that, the outage for IoT network persists. Apparently, if $\eta_s$, $\eta_t$ are proportional to $\eta$, the diversity order of the IoT network reduces to zero. Here, the tightness of analytical lower bound OP lies over medium to SNR region only. However, the adaptive choice of $\mu$ lead to significant improvement in OP even with zero diversity order. Although the choice of fixed $\mu$ results in lower OP of IoT network at low SNR regime (as compared to adaptive $\mu$ with OP unity), the desired QoS of satellite network may not be guaranteed.    

\section{Conclusion}\label{con}
We have analyzed the outage performance of an OSTN where the primary satellite-to-terrestrial communications are enabled by a secondary IoT network in the presence of combined interference from extra-terrestrial and terrestrial sources. We derived closed-form lower bound OP for both the satellite and IoT network under opportunistic IoT transmitter selection strategy. We further derived the asymptotic OP for the two networks to reveal their achievable diversity order in the presence of interference. We observe that the adaptive choice of power splitting factor significantly improves the OP of IoT network while guaranteeing the QoS of satellite network.  


\appendices
\section{}\label{appA}
To derive (\ref{cmqr}), first, we approximate the term $\hat{\Lambda}_{ac_{k}}=\frac{{\Lambda}_{ac_{k}}}{W_c+1}$ in (\ref{snrs}) as $\hat{\Lambda}_{ac_{k}}\approx\frac{{\Lambda}_{ac_{k}}}{W_c}$ under high interference power to upper bound the exact SINR $\Lambda_{ac_{k}b}$. Then, we apply the bound $\frac{XY}{X+Y}\leq\min(X,Y)$ to re-express (\ref{snrs}) as  
${\Lambda}_{ac_{k}b}\leq\frac{\mu}{(1-\mu)+1/\min({\hat{\Lambda}_{ac_{k}},\Lambda_{c_{k}b}})}.$
Substituting this SINR in (\ref{cmqr}), we get 
\begin{align}\label{hsi}
\tilde{\mathcal{P}}^{\textmd{sat}}_{\textmd{out}}(\mathcal{R}_{p})&=\mathbb{E}\{\sum_{n=0}^{K} \binom{K}{n}(-1)^n[\overline{F}_{\hat{\Lambda}_{ac_k}}(\tilde{\gamma}_p|w)]^n\\\nonumber
&\times[\overline{F}_{\Lambda_{c_kb}}(\tilde{\gamma}_p)]^n\},
\end{align}
where $\overline{F}_{X}(\cdot)=1-F_{X}(\cdot)$. 

To proceed further, we require the pdf of $W_c$ (i.e., the combined extra-terrestrial and terrestrial interference).
\begin{lemma}\label{lem1}
	The pdf of $W_c$ is given as
	\begin{align}\label{wc}
	f_{W_c}(w)&=\widetilde{\sum}\frac{\Xi(M_1)}{\eta^\Lambda_s}\left(\frac{1}{\Omega_t\eta_t}\right)^{M_2}\frac{\Phi(M_2,\Lambda)}{\Gamma(M_2)}\\\nonumber
	&\times w^{\Lambda+M_2-1}\textmd{e}^{-\frac{w}{\Omega_t\eta_t}}{_1F}_1\left(\Lambda;M_2+\Lambda;-\Theta_{s,s}w\right),
	\end{align}
	where ${_1F}_1(\cdot;\cdot;\cdot)$ is the confluent hypergeometric function \cite[eq. 9.210]{grad}.
\end{lemma}
\begin{IEEEproof}
	As $W_c= W_s+W_t$, the pdf of $W_c$ can be evaluated as the convolution of independent and non-identically distributed (i.ni.d.) pdfs of $W_s$ and $W_t$ as
	\begin{align}\label{pdfi}
	f_{W_c}(w)&=\int_{0}^{w}f_{W_s}(x)f_{W_t}(w-x)dx.
	\end{align}
	To calculate (\ref{pdfi}), the pdf of $W_s$ (i.e., sum of i.i.d. and equal power extra-terrestrial shadowed-Rician interferers) can be given as \cite{pku}
	$f_{W_s}(x)$$=$$ \sum_{i_{1}=0}^{m_{s}-1}\cdots \sum_{i_{M_1}=0}^{m_{s}-1}\frac{\Xi(M_1)}{(\eta_{s})^{\Lambda}}x^{\Lambda-1} \textmd{e}^{-\tilde{\Theta}_{s,s}x}$, where $\Xi(M_1)=\alpha^{M_1}_s\prod_{\kappa=1}^{M_1}\zeta(i_{\kappa})\prod_{j=1}^{M_1-1}\Phi(\sum_{l=1}^{j}i_{l}+j,i_{j+1}+1)$, $\Lambda=\sum_{\kappa=1}^{M_1}i_{\kappa}+M_1$. 
	Further, the pdf of $W_t$ (i.e., sum of i.i.d. and equal power terrestrial Rayleigh interferers) is given as \cite{kangc}
	$f_{W_t}(x)$$=$$\Big(\frac{1}{\Omega_t\eta_t}\Big)^{M_2}\frac{x^{M_2-1}}{\Gamma(M_2)}\textmd{e}^{-\frac{x}{\Omega_t\eta_t}}$.
	After invoking these pdfs into (\ref{pdfi}), we apply \cite[eq. 3.383]{grad} to get (\ref{wc}). 
	\end{IEEEproof}

Furthermore, in (\ref{hsi}), the cdf ${F}_{\hat{\Lambda}_{ac_k}}(x|w)$ can be obtained with the aid of shadowed-Rician pdf as 
\begin{align}\label{pdflsz}
{F}_{\hat{\Lambda}_{ac_k}}(x|w)&=1-\alpha_{c}\sum_{l=0}^{m_{ac}-1}\frac{\zeta(l)}{(\eta_{a})^{l+1}}\sum_{p=0}^{l} \frac{l!}{p!}\Theta^{-(l+1-p)}_{c,a}\\\nonumber
&\times (xw)^{p}\textmd{e}^{-\Theta_{c,a}xw}.
\end{align}
Also, we can write the cdf ${F}_{{\Lambda}_{c_kb}}(x)=1-\textmd{e}^{-\frac{x}{\Omega_{cb}\eta_c}}$. Eventually, making use of the aforementioned pdfs in (\ref{hsi}) and applying the multinomial expansion \cite{pkss} along with the identity \cite[eq. 07.20.26.0006.01]{wolf} for ${_1F}_1(\cdot;\cdot;\cdot)$, and evaluating the resulting expression using \cite[eq. 7.813]{grad}, one can attain (\ref{ajsl23}). 
 


\end{document}